\documentclass[preprint, showpacs, aps, draft]{revtex4}
\usepackage{bm}

\begin{document}
\title{Prior information: how to circumvent the standard joint-measurement
uncertainty relation} 
\author{Michael J. W. Hall}
\affiliation{Theoretical Physics, IAS,\\ Australian National
University,\\
Canberra ACT 0200, Australia}

\begin{abstract}
The principle of complementarity is quantified in two ways: by a universal
uncertainty relation valid for arbitrary joint estimates of any two
observables from a given measurement setup, and by a general
uncertainty relation valid for the {\it optimal} estimates of the same two
observables when the state of the 
system prior to measurement is known.
A formula is given for the optimal estimate of any
given observable, based on arbitrary measurement data and prior information
about the
state of the system, which generalises and provides a more robust
interpretation of previous formulas for ``local expectations'' and
``weak values'' of quantum
observables.   As an example, the 
canonical joint measurement of position $X$ and momentum $P$ 
corresponds to measuring the commuting
operators $X_J=X+X'$, $P_J=P-P'$, where
the primed variables refer to an auxilary system in a
minimum-uncertainty state.  It is well known that   
$\Delta X_J\,\Delta P_J\geq\hbar$. 
Here it is shown that given the {\it same}
physical experimental setup, and knowledge of the system density
operator prior to measurement, one can make improved joint estimates
$X_{opt}$ and $P_{opt}$ of $X$ and $P$.  These improved estimates are not
only statistically closer to $X$ and $P$:  they satisfy $\Delta
X_{opt}\,\Delta P_{opt}\geq\hbar/4$, where equality can be achieved in certain
cases.  Thus one
can do up to four times better than the standard lower bound (where the latter 
corresponds to the limit of {\it no}
prior information). Other applications include the  
heterodyne detection of orthogonal quadratures of a single-mode optical 
field, and joint measurements based on Einstein-Podolsky-Rosen
correlations. 
\end{abstract}
\pacs{03.65.Ta}
\maketitle
\newpage

\section{Introduction}

At least four generic types of uncertainty principle can be distinguished in 
quantum theory:\\
(i) {\it State preparation:} the quantum description of a physical system cannot 
simultaneously assign definite values to all observables;\\
(ii) {\it Overlap:} different physical states cannot in general be unambiguously 
distinguished by measurement;\\
(iii) {\it Disturbance:} measurement of one observable necessarily ``disturbs'' 
other observables;\\
(iv) {\it Complementarity:} the experimental arrangements for accurately 
defining/measuring different observables are in general physically incompatible.

These principles are all negative in content, corresponding to limits on what is 
possible in quantum mechanics. These limits are quantified via associated 
uncertainty relations.  As the literature on such uncertainty relations is 
extensive, only a few general remarks and indicative references will be given 
here to set the context for this paper.  

The ``state preparation'' uncertainty principle is the best known, and places 
limitations on classical notions of prior knowledge (and hence predictability). 
The 
corresponding uncertainty relations are generally expressed in terms of the 
spreads of the probability distributions of different observables, the 
prototypical example being the textbook  inequality 
\begin{equation}\label{un1}
\Delta X\,\Delta P\geq\hbar/2
\end{equation}
for the rms spreads of position and momentum.

The ``overlap'' uncertainty principle corresponds to the existence of 
non-orthogonal states, and underlies the 
semi-classical notion that quantum states 
occupy a phase space area of at least $2\pi\hbar$.  It also separates quantum 
parameter estimation from its classical counterpart. A corresponding uncertainty 
relation is the parameter estimation bound
\begin{equation}\label{un2}
 \delta X\,\Delta P\geq \hbar/2  
\end{equation}
where $\delta X$ is a measure of the error in any (covariant) estimate of the 
amount by which a state has been displaced in position, and $\Delta P$ is the 
rms momentum spread of the state \cite{helstrom, holevo,caves}.
 
The ``disturbance'' uncertainty principle is connected to early statements by 
Heisenberg such as `every subsequent observation of the position will alter the 
momentum by an unknown and undeterminable amount' \cite{heisp20}.  
Investigation of this principle has proceeded by examining the distribution
of one observable both before and after the measurement of another
observable, and attempting to relate the disturbance of the distribution to the 
accuracy of the measurement \cite{aharonov, appleby, dariano}. However, recent 
work by Ozawa
\cite{ozawapla, ozawa} shows that the momentum disturbance $\eta(P)$ due to a 
position measurement having inaccuracy $\epsilon(X)$ can in fact satisfy 
$\epsilon(X)\eta(P)=0$.  Hence this principle needs to be formulated more 
carefully, presumably in relation to valid uncertainty relations such as 
\cite{ozawapla, ozawa} 
\begin{equation}\label{un3}
\epsilon(X)\,\eta(P)+\epsilon(X)\Delta P +\Delta X\,\eta(P)\geq\hbar/2 .  
\end{equation}

Finally, the fourth uncertainty principle above arises from Bohr's notion of 
{\it complementarity} \cite{bohr}, and restricts the degree to which {\it joint} 
information about observables can be obtained from a single experimental setup. 
However, previous formulations of corresponding uncertainty relations have only 
been given for special cases \cite{appleby, ak, wootters, ag, martens,whichway, 
muynck,trif}.   The most general of these are
the Arthurs-Kelly type \cite{ appleby, ak, ag,trif}, restricted to ``universally 
unbiased'' joint measurements; and the Martens-deMuynck type 
\cite{martens,muynck}, restricted to ``non-ideal'' joint measurements.  For 
example, if a measurement apparatus simultaneously outputs two values $X_J$ and 
$P_J$, that are on average equal to the averages of $X$ and $P$ (for {\it all} 
input states), then \cite{ appleby, ak, ag} \begin{equation}\label{un4}
 \Delta X_J\Delta P_J \geq\hbar. 
\end{equation}
The need to find {\it general} uncertainty relations quantifying 
complementarity, {\it not} subject to any 
restrictions on measurement, forms the 
subject of this paper.

To proceed, one first clearly needs to generalise what is meant by a 
``joint measurement''.  For example, neither ``universally unbiased'' or 
``non-ideal''
joint measurements include experiments that are adapted in some way to 
particular subclasses of states.  Yet Bohr defended complementarity against a 
number of thought experiments of this type \cite{bohr}, including the famous 
Einstein-Podolsky-Rosen (EPR) paradox \cite{bohr, epr, bohrepr}.  In the latter 
case a joint measurement of $X$ and $P$ arises via simultaneous measurement of 
$X$ and $P'$, where $P'$ refers to the momentum of a (correlated)
auxilary 
system.  Such a joint measurement does not satisfy either of the ``universal 
unbiasedness'' or ``non-ideal'' restrictions mentioned  above.

Indeed, in trying to place fundamental limits on the information which can 
simultaneously be gained about two complementary observables, one must consider 
{\it any and all} experimental setups, without restriction. The simplest and 
most general possible approach will therefore be taken in this paper:  {\it any 
measurement is considered to provide a joint measurement of any two 
observables}.  The corresponding logic is that (i) the result of a given 
measurement provides information; (ii) this information can be used to make an 
{\it estimate} of any given observable; and (iii) one may look for universal 
uncertainty relations associated with such estimates.  

This approach solves the problem of what constitutes a joint measurement in a 
very general way ({\it all} measurements are permitted).  However, there
are still two possible strategies that may be followed to obtain
general joint-measurement uncertainty relations.  The first of 
these is simply to seek
uncertainty relations which hold for {\it any} estimates, good or bad, of the
observables.  The second strategy is to throw away all the bad estimates,
and only seek uncertainty
relations for estimates that make the best possible use of
any prior information (after all, why make a particular estimate 
if the information
is available to make a
better one?).  
Both strategies will be followed in this paper, and corresponding
joint-measurement uncertainty relations are given in Secs.~III and IV.

Note that the strategy of making the best use of any available 
prior information is of some 
interest in its own right, quite aside from joint measurements.  
For example, a measurement of position does not by itself 
provide a very useful basis for estimating energy. 
However, combining the measurement result with any information available
about the system {\it before} measurement 
(eg, its average momentum, or its quantum state, or its
entanglement with an auxilary system) can lead to a
significantly improved estimate.
More generally, prior information helps the experimenter place the 
detector to minimise null outcomes, and the quantum communications engineer to 
optimise the receiver.  As emphasised by Trifonov et al.~\cite{trif}, even the 
``universally unbiased'' bound in Eq.~(\ref{un4}) is achieved only by choosing 
the experimental setup in dependence on prior 
information about the system to be 
measured (the ``balance'' parameter $b=\hbar\Delta X/\Delta P$ 
in \cite{ak}, and 
the full polarisation state in \cite{trif}).  

In Sec.~II a general formula is given for the best possible estimate of an 
observable, based on an arbitrary measurement and prior information about the 
state of the system.  This formula is related to and 
generalises expressions for 
``local expectation values'' \cite{wan, holland,cohen} and ``weak values'' 
\cite{weak, weakpra} of quantum observables.  
The best possible estimate is also 
determined for the case in which there is {\it no} prior information available. 
Examples are given for general energy estimation, and for estimation of the  
quadratures of a single-mode field using optical heterodyne detection. In the 
latter case the best possible estimates are related to the gradient of the 
Husimi Q-function.

In Sec.~III a geometric uncertainty relation is given for the optimal estimates 
of any two observables from arbitrary measurement data, assuming that the state 
prior to measurement is known.  This uncertainty relation implies a trade-off 
between the {\it dispersions} of the estimates (i.e., the spreads of the 
corresponding distributions), and the {\it inaccuracies} of 
the estimates (i.e., the 
degree to which the estimates successfully 
mimic the corresponding observables).  A 
universal lower bound for the inaccuracy of any (possibly non-optimal)
estimate is also given.  For the 
case of heterodyne detection two further inequalities are derived, applying to 
the dispersions and to the inaccuracies of the estimates respectively.  It is 
also  shown that the optimal estimates resulting from a canonical joint 
measurement of position and momentum, on a known state, 
satisfy an uncertainty relation with a 
lower bound 1/4 of that in Eq.~(\ref{un4}).  

In Sec.~IV a {\it universal} joint-measurement uncertainty relation is derived, 
valid for {\it any} 
estimates (optimal or otherwise) of two observables from a given 
experimental setup.  The derivation shares a formal link with Ozawa's proof of 
Eq.~(\ref{un3}) \cite{ozawapla, ozawa}, and modification of the derivation leads 
to stronger uncertainty relations such as Eq.~(\ref{un4}) for the special case 
of universally unbiased measurements. Results are applied to a discussion of the 
above-mentioned EPR paradox \cite{epr}, and to quadrature estimation based on 
prior information about the averages of certain observables.

Some conclusions are given in Sec.~V.

\section{Making the best possible estimate}

Consider an arbitrary measurement ${\cal M}$ having possible results $\{m\}$, 
and with statistics given by
\begin{equation} \label{prob}
 p(m|\rho) = {\rm tr}[\rho M_m] 
\end{equation}
for a system described by density operator $\rho$.  Since the probabilities must 
be positive and sum to unity, the operators $\{ M_m\}$ must be positive and sum 
to the unit operator, and hence form a {\it probability operator measure} (POM) 
\cite{helstrom, holevo,lahti}.  In the interests of generality, no further 
restrictions or specific measurement models are assumed.   

A notation is adopted whereby a measurement, its
corresponding POM, and the corresponding observable quantity, are all
denoted by the same scripted character, eg, ${\cal M}$.  Any Hermitian
operators associated with the measurement will be denoted via related upper-case Roman characters, eg, $M_m$.

In some cases ${\cal M}$ may be equivalently described by a Hermitian 
operator $M$.  In such a case $M_m$ is just the projection onto the eigenspace 
associated with eigenvalue $m$ of $M$, and $M=\sum_m mM_m$.  
If $M$ is non-degenerate with eigenkets 
$\{ |m\rangle\}$, and the system is described by a pure state $\rho = 
|\psi\rangle\langle\psi |$, Eq.~(\ref{prob}) reduces to the familiar expression 
$p(m|\psi) = |\langle m|\psi\rangle |^2$.  
As is well known, however, there are many non-trivial measurements that 
are not equivalent to some
Hermitian operator acting on the Hilbert space of the system 
\cite{helstrom,holevo,lahti}.

As discussed in the Introduction, it may often be desirable to make an
estimate of some observable based on the result of a given
measurement and any available prior information.  
This Section is therefore concerned with answering the following question:  
{\it for a quantum system described by density 
operator} $\rho$, {\it what is the best possible estimate one can make of some 
observable}, ${\cal A}$, {\it from a measurement of} ${\cal M}$ {\it with 
result} $m$?  

\subsection{Using prior information: a special case}

It is convenient, for the purpose of introducing the
necessary concepts, to first consider the above question in the 
special case that ${\cal M}$ and ${\cal A}$ correspond to respective 
Hermitian operators $M$ and $A$.  
This case was also considered briefly in Ref.~\cite{halleur}.

In particular, suppose that one seeks the best possible estimate of 
$A$ from the measurement of a Hermitian operator $M$ having
eigenkets $\{|m\rangle\}$, where for simplicity it will be assumed that the system is in a known {\it pure} state $|\psi\rangle$ prior to measurement.  It follows that any estimate $f(m)$ of $A$ from measurement
result $M=m$ is equivalent to measurement of the Hermitian operator
$f(M)=\sum_m f(m) |m\rangle\langle m|$.
One may therefore represent the estimate as
\[ f(M) = A + N_f , \]
i.e., as the sum of the operator to be estimated, $A$, and a
``noise operator'', $N_f$ \cite{appleby,ozawapla,ag}.

Now, the best possible estimate will of course depend on the
criterion of optimality used to define ``best possible''. One obvious criterion is that the noise should be ``small'' on average, i.e., the quantity
\begin{equation} \label{noisy}
 \langle \psi|N_f^2|\psi\rangle =
\langle\psi| [f(M)-A]^2|\psi\rangle = D_\psi(f(M),A)^2
\end{equation} 
should be small.  Here 
$D_\psi(A,B):=\langle (A-B)^2\rangle^{1/2}$ denotes the {\it statistical 
deviation} between Hermitian operators $A$ and $B$ (see also Appendix).  
The best possible estimate is therefore defined as
corresponding to {\it the choice of $f$ that minimises the statistical
deviation between the observable and its estimate}.  

To determine this best possible estimate, note that
Eq.~(\ref{noisy}) can be rewritten as
\begin{eqnarray}
D_\psi(f(M),A)^2 
& = & \langle A^2\rangle + \sum_m |\langle m|\psi\rangle|^2 f(m)^2 
- \sum_m f(m) [\langle
\psi|m\rangle\langle m|A|\psi\rangle + c.c.] \nonumber\\
& = & \langle A^2\rangle - \sum_m |\langle m|\psi\rangle|^2 \left[ {\rm
Re}\, \frac{\langle m|A|\psi\rangle}{\langle m|\psi\rangle}\right]^2
\nonumber\\ \label{devpure}
& &~~~+ \sum_m |\langle m|\psi\rangle|^2 \left[ f(m) - {\rm Re}\,
\frac{\langle m|A|\psi\rangle}{\langle m|\psi\rangle}\right]^2 . 
\end{eqnarray}
Only the last term depends on the estimate, and is nonnegative. Hence the
minimum possible statistical deviation or ``noise'' corresponds to
the choice \cite{halleur}
\begin{equation}
\label{bestpure}
f(m)=\tilde{A}_{\rm opt}(m|\psi):= {\rm Re}\, \frac{\langle
m|A|\psi\rangle}{\langle
m|\psi\rangle} .
\end{equation}
A tilde is used to distinguish this quantity from an operator.

Thus, when statistical deviation is used as the criterion of 
optimality, the optimal estimate of $A$, from measurement
result $M=m$ on state $|\psi\rangle$, is given by 
$\tilde{A}_{\rm opt}(m|\psi)$ in Eq.~(\ref{bestpure}).  It is only possible to
make this estimate when the appropriate prior information - 
the state prior to measurement - is known.  The case where no prior
information is available is considered in Sec.~II.C below.

The formula on the righthand side of Eq.~(\ref{bestpure}) has in fact appeared
previously in the literature in a variety of other contexts:  as the ``local
expectation value'' of the operator $A$ relative to $M$ for state
$|\psi\rangle$
\cite{wan,holland, cohen}; as the ``weak value'' of the operator $A$
relative to
pre-selected state $|\psi\rangle$ and post-selected state $|m\rangle$
\cite{weak, weakpra, weakwiseman, lars1}; and as the ``classical component''
of $A$
with respect to $M$ for state $|\psi\rangle$ \cite{halleur, hallfish,
luis}.
However, only the above
``estimation'' context appears to provide a robust interpretation. 

For example,
the expression in Eq.~(\ref{bestpure}) can be negative for a positive
operator
$A$, which undermines its interpretation as either a ``value'' or a
``classical'' component of $A$.  In contrast, the
fact that the best possible estimate of some positive observable, from
the measurement of a second incompatible observable,
can be negative on occasion 
merely provides a nice signature of the difference
between quantum and classical estimation theory \cite{classical}
(at least in the case
where statistical deviation is used as the sole criterion of optimality).
While one could of course restrict attention to
estimates that fall within the eigenvalue range of $A$, the estimate in
Eq.~(\ref{bestpure}) still remains of fundamental interest in providing an
absolute lower bound for the statistical deviation of {\it any} estimate. It should be noted
that, in any case, all
examples considered in this paper satisfy this restriction (with the exception of Eq.~(\ref{qx}) in
Sec.~II.D). 

\subsection{Using prior information: the general case}

The question posed at the beginning of this section, of how to
determine the  
best possible estimate of an observable ${\cal A}$ from measurement of a
{\it general} POM observable ${\cal M}$ on a system described by a known
density operator $\rho$, may now be addressed.  Clearly, 
it is first necessary to suitably generalise in some way the 
criterion of optimality used in the previous section.

For the case where ${\cal A}$
corresponds to some Hermitian operator $A$, the generalisation of statistical deviation turns out to be 
quite straightforward.  In particular, as discussed in the Appendix, the natural definition of the 
statistical deviation between a Hermitian operator $A$ and a POM
observable ${\cal B}\equiv\{B_b\}$, for a given state $\rho$, is
\begin{equation} \label{natdev} 
 D_\rho (A, {\cal B})^2 := \sum_b {\rm tr}[(A-b)\rho (A-b)B_b] .
\end{equation} 
Note that this reduces to $D_\psi(A,B)$ in Eq.~(\ref{noisy}) 
when ${\cal B}$ corresponds to some Hermitian
operator $B$.  As shown in the Appendix, the derivation in 
Eq.~(\ref{devpure}) is easily generalised to give the optimal estimate 
\begin{equation} \label{best}
f(m)=\tilde{A}_{\rm opt}(m|\rho) := \frac{{\rm tr}[\rho(M_m A + AM_m)]}{2\,
{\rm tr}[\rho M_m]} 
\end{equation}
of $A$ from measurement result ${\cal M}=m$.  The 
case where ${\cal A}$ does {\it not} correspond to a Hermitian
operator is also discussed in the Appendix.

Eq.~(\ref{best}) clearly generalises Eq.~(\ref{bestpure}), and has 
several properties worth noting.  First, it
follows via Eq.~(\ref{prob}) that the optimal 
estimate is always {\it unbiased}, i.e., 
\begin{equation} \label{av}
\sum_m p(m|\rho)\, \tilde{A}_{\rm opt}(m|\rho) = {\rm tr}[\rho A] =\langle A\rangle .
\end{equation} 
Second, if the system is initially in some eigenstate $|a\rangle$ of $A$, then 
$\tilde{A}_{\rm opt}(m|\rho) \equiv a$, independently of the actual measurement result.  
Third, if ${\cal M}$ corresponds to an {\it ideal} measurement of a Hermitian 
operator which commutes with $A$, one has the classical repeatability property
\begin{equation} \label{repeat}
\tilde{A}_{\rm opt}(m|\overline{\rho})= \tilde{A}_{\rm opt}(m|\rho) ,
\end{equation}
where $\overline{\rho} := \sum_{m'} M_{m'} \rho M_{m'}$ describes the 
post-measurement ensemble.  

It is convenient to denote the physical 
observable associated with the optimal estimate in Eq.~(\ref{best}) by
${\cal A}_{\rm opt}$.  Measurement of ${\cal A}_{\rm opt}$ is 
carried out by measuring ${\cal M}$, and for result 
$m$ attributing the outcome 
$\tilde{A}_{\rm opt}(m|\rho)$ to ${\cal A}_{\rm opt}$. One 
may refer to ${\cal A}_{\rm opt}$ as the {\it compatible} component of $A$ with 
respect to ${\cal M}$.  Note  from Eq.~(\ref{best}) that compatible components 
form a linear algebra, i.e.,
\[ ({\cal A+\lambda B})_{\rm opt} = {\cal A}_{\rm opt} + \lambda{\cal B}_{\rm opt}.\]

\subsection{No prior information}

Consider now the case where there is {\it no} information available
about the state of the 
system prior to measurement.  The statistical deviation therefore cannot be 
calculated, nor the estimates in Eqs.~(\ref{bestpure}) and (\ref{best}).
The best possible estimate of $A$ from some measurement 
${\cal M}$ must instead be defined via some state-{\it independent} 
criterion of optimality.

One suitable criterion is provided by a generalisation of the 
Hilbert-Schmidt distance $d(A,B)^2={\rm tr}[(A-B)^2]$ 
between two Hermitian operators. Such a generalisation, 
$d(A,{\cal M})$, for a Hermitian operator $A$ and a POM 
observable ${\cal M}$, has recently been given \cite{hallalg}.  
The estimate ``closest'' to $A$, in the sense of minimising 
this distance, follows directly as (see Appendix)
\begin{equation} \label{bestnoinf}
\tilde{A}^0_{\rm opt}(m) := {\rm tr}[AM_m]/{\rm tr}[M_m] .
\end{equation}
The physical observable corresponding to this estimate will be denoted by ${\cal 
A}^0_{\rm opt}$. Note from Eq.~(\ref{best}) that 
${\cal A}^0_{\rm opt} \equiv {\cal A}_{\rm opt}$ in 
the case that $\rho$ is a maximally mixed state, i.e., $\rho\sim{\bf 1}$.  

The estimate in Eq.~(\ref{bestnoinf}) is typically biased - after all, there is 
no prior information available about $\langle A\rangle$ to feed into such an 
estimate.  However, depending on the relationship between $A$ and ${\cal M}$, it 
is possible for the estimate to be universally unbiased, as will be seen for 
heterodyne detection in Sec.~II.D below.  Further, in cases where the estimate 
is only {\it linearly} biased, it is possible to trade ``distance'' for 
``bias''.  For example, if
\begin{equation} \label{linbias}
\sum_m M_m\, {\rm tr}[AM_m]/{\rm tr}[M_m] = A+r 
\end{equation}
for some constant $r$, then a universally unbiased estimate is obtained by 
replacing $A^0_{\rm opt}(m)$ with $A^0_{\rm opt}(m) - r $.  

As a more general example, consider the estimate of the spin, ${\bf S}= 
\hbar{\bf \sigma}/2$, of a spin-1/2 particle, from a measurement result ${\bf 
m}$ corresponding to a general POM $\{M_{\bf m} = q_{\bf m}(1+ {\bf \sigma . 
m})\}$. Here ${\bf m}$ ranges over some subset $R$ of the Bloch ball, and 
$\{q_{\bf m}\}$ is any probability distribution on $R$ satisfying $\sum_{\bf m} 
q_{\bf m}{\bf m} =0$.  The  best possible estimate of ${\bf S}$ from result 
${\bf m}$ follows from Eq.~(\ref{bestnoinf}) as the linearly biased estimate 
$\hbar {\bf m}/2$.  The associated universally unbiased estimate is $\hbar 
\Lambda^{-1}{\bf m}/2$, where $\Lambda$ denotes the matrix $\sum_{\bf m} q_{\bf 
m}{\bf m\, m}^T$ (note that the inverse exists providing $R$ contains three 
linearly independent members).  A similar result holds on general Hilbert spaces  
for trace-class $M_m$ and $A$, with the components of ${\bf \sigma}$ replaced by 
a linearly independent basis set of trace-free Hermitian operators. 

\subsection{Example: energy estimation}

Making the best possible estimate of energy from the measurement of various 
observables is considered here, to indicate the types of expressions that can 
arise.

First, for a particle with Hamiltonian operator $H$, consider the case where all 
that is known about the system is that it is in thermal equilibrium 
corresponding to temperature $T$.  The particle is therefore described by the 
density operator proportional to $e^{-\beta H}$, where $\beta=1/(kT)$.  It 
follows from Eq.~(\ref{best}) that the optimal estimate of the energy of the 
system, from measurement result ${\cal M}=m$, is given by
\begin{equation} \label{thermal}
\tilde{E}_{\rm opt}(m|T) = -(\partial/\partial\beta) \ln {\rm tr}[e^{-\beta H}M_m] . 
\end{equation}
Thus ${\rm tr}[e^{-\beta H}M_m]$ is a kind of generalised partition function.  
For the particular example of a position measurement, on a 1-dimensional 
harmonic oscillator of mass $m$ and frequency $\omega$, one obtains the 
quadratic estimate
\[  \tilde{E}_{\rm opt}
(x|T) = A_T + B_T\,x^2 ,\]
where $A_T=(1/2)\hbar\omega\coth (\beta\hbar\omega)$ and $B_T= (1/2)m 
\omega^2{\rm sech}^2(\beta\hbar\omega/2)$.  Note that in the zero-temperature 
limit this estimate reduces to the groundstate energy $\hbar \omega /2$, 
independently of the actual measurement result $x$. In the classical limit
$\hbar\rightarrow 0$ the estimate reduces to $(1/2)kT+(1/2)m\omega^2x^2$, i.e., to the sum of the average kinetic energy and the potential energy (this result holds more generally).

Second, for a particle with Hamiltonian $H=P^2/(2m)+V(X)$ in a known {\it pure} 
state $|\psi\rangle$, the best possible estimate of energy from a measurement of 
position $X$ follows from Eq.~(\ref{bestpure}) as 
\begin{equation} \label{qx}
\tilde{E}_{\rm opt}(x|\psi) = |\nabla S|^2/(2m) + V(x) + Q(x),
\end{equation}
where $\langle x|\psi\rangle=R\exp(iS/\hbar)$, and $Q(x)= -
\hbar^2/(2m)\nabla^2R/R$ is the so-called ``quantum potential'' \cite{holland}. 
Note that $Q(x)$ arises here in the context of the best possible estimate of the 
kinetic energy [i.e., the possibly negative quantity $|\nabla S|^2/(2m) + 
Q(x)$], with no relation to a real potential energy.

Third and finally, consider a single-mode optical field with annihilation 
operator $a$ and Hamiltonian $H=\hbar\omega a^\dagger a$.  An inefficient 
measurement of photon number, via a photodetector having quantum efficiency 
$\eta$, corresponds to the POM $\{ M_m(\eta)\}$ with number-state expansion 
$M_m(\eta) = \sum_r |m+r\rangle\langle m+r|\,\, {}^{m+r}\!C_r\eta^m (1-\eta)^r$ 
\cite{muynck,lahti}. If there is {\it no} prior information about the state of 
the field prior to measurement, the best possible estimate of the energy of the 
field then follows from Eq.~(\ref{bestnoinf}) as
\begin{equation} \label{ineff}
\tilde{E}^0_{\rm opt}(m)= \hbar\omega[(m+1)/\eta -1] , 
\end{equation}
using the identity $\sum_r {}^{m+r}\!C_rx^r = (1-x)^{-m-1}$.  This estimate is 
linearly biased [with $r = 1/\eta -1$ in Eq.~(\ref{linbias})], with the 
associated univerally unbiased estimate given by $\hbar\omega m/\eta$.

\subsection{Example: heterodyne detection}

For a single-mode optical field with annihilation operator $a$, the quadrature 
observables $X_1=(a+a^\dagger)/2$, $X_2=(a-a^\dagger)/2i$ have commutator 
$[X_1,X_2]=i/2$, and hence are analogous to the position and momentum 
observables of a quantum particle (with $\hbar$ replaced by 1/2).  In 
particular, $X_1$ and $X_2$ cannot be measured simultaneously to an arbitrary 
accuracy.  

However, in optical heterodyne detection, one introduces an auxilary imageband 
field with annihilation operator $b$, and simultaneously measures the real and 
imaginary parts of the operator $a+b^\dagger$ \cite{yuen, shapiro, cd}, i.e., 
one measures the commuting observables
\begin{equation} \label{jointquad}
  {X}_{1,J}=X_1+Y_1,~~~~~~~ {X}_{2,J}=X_2-Y_2 , 
\end{equation}
where $Y_1$ and $Y_2$ denote the corresponding quadratures of the imageband 
field.  This may be interpreted as corresponding to an approximate joint 
measurement of $X_1$ and $X_2$, subject to imageband noise. 

Clearly this joint measurement is formally equivalent to the canonical joint 
measurement of position $X$ and momentum $P$ of a quantum particle, referred to 
in the Abstract, where one introduces an auxilary particle with corresponding 
observables $X'$ and $P'$, and simultaneously measures 
\begin{equation} \label{xp}
X_J=X+X',\,\,\,\,\,\,\,\,\,\,\,\,  P_J=P-P'.
\end{equation}  
This formal equivalence allows one to map results from one context to the other.

For the general case of an uncorrelated imageband field described by density 
operator $\rho_i$, the measurement statistics of heterodyne detection correspond 
to a continuous POM $\{M_\alpha\}$, with
$M_\alpha = \pi^{-1} D(\alpha)\rho'_iD(\alpha)^\dagger$ \cite{hallfuss}. 
Here $\alpha$ denotes the complex eigenvalue $\alpha_1+i\alpha_2$ of 
$a+b^\dagger$, $D(\alpha)$ denotes the Glauber displacement operator $\exp 
(\alpha a^\dagger-\alpha^*a)$, and $\rho'_i$ is defined by 
\[ \rho'_i := \sum_{m,n} |m\rangle_a \,{}_a\langle n| (-1)^{m+n} 
{}_b\langle m|\rho_i|n\rangle_b^*  , \]
where the subscripts $a$ and $b$ refer to number states of the signal and 
imageband fields respectively.  

For simplicity, attention will be restricted in what follows to the case of a 
vacuum-state imageband field.  For this case $\rho'_i = \rho_i = 
|0\rangle\langle 0|$, and hence the measurement is described by the well known 
coherent-state POM \cite{helstrom,holevo,lahti,cd}, with 
\[ M_\alpha=\pi^{-1}|\alpha\rangle\langle\alpha|,\]
and associated 
measurement statistics given by the Husimi Q-function 
\begin{equation} \label{q}
Q(\alpha)=\pi^{-1}\langle\alpha|\rho|\alpha\rangle .
\end{equation}

Now, suppose first that there is {\it no} prior information available about the 
state of the field. The best possible estimate of the quadrature $X_1$, from 
measurement result $\alpha$, then follows from Eq.~(\ref{bestnoinf}) as the 
estimate
\begin{equation} \label{x0}
\tilde{X}^0_{1,{\rm opt}}(\alpha)  =  \langle\alpha|X_1|\alpha\rangle 
/\langle\alpha|\alpha\rangle = \alpha_1 .
\end{equation}
Similarly, the best possible estimate of $X_2$ in the case of no prior 
information is given by
\begin{equation} \label{p0} 
\tilde{X}^0_{2,{\rm opt}}(\alpha) = \alpha_2 .
\end{equation}
Thus the best possible estimates are directly given by the measurement result 
$\alpha$, i.e., they are {\it equivalent} to measurement of ${X}_{1,J}$ and 
${X}_{2,J}$ in Eq.~(\ref{jointquad}).  More generally, the best possible estimate 
of a general Hermitian observable $f(a,a^\dagger)$, when no prior information is 
available, follows from Eq.~(\ref{bestnoinf}) as $f^{(n)}(\alpha, \alpha^*)$, 
where $f^{(n)}$ denotes the normally-ordered form of $f$.

The situation changes markedly when prior information about the state of the 
system {\it is} available.  In particular, the best possible estimate of $X_1$ 
for a measurement on a known state $\rho$ follows via Eq.~(\ref{best}) as
\begin{eqnarray}
\tilde{X}_{1,{\rm opt}}(\alpha|\rho) & = & \langle\alpha |X_1
\rho+\rho X_1 |\alpha \rangle/ 
\langle\alpha|\rho|\alpha\rangle /2 \nonumber\\
& = & \frac{1}{2}\,{\rm Re}\left\{ \alpha + \langle\alpha|a\rho| \alpha\rangle 
/\langle\alpha|\rho|\alpha\rangle \right\} . \label{bestcoh1}
\end{eqnarray}
Thus the direct ``no prior information'' estimate, $\alpha_1 = {\rm Re}\,\alpha$ 
in Eq.~(\ref{x0}), provides only {\it half} of the input to the more general 
estimate of $X_1$.  The other half depends on the state, and is typically a 
highly nonlinear function of both $\alpha_1$ and $\alpha_2$.  One has a similar 
estimate
\begin{equation} \label{bestcoh2}
\tilde{X}_{2,{\rm opt}}(\alpha|\rho) = \frac{1}{2}\,{\rm Im}\left\{ \alpha + 
\langle\alpha|a\rho| \alpha\rangle /\langle\alpha|\rho|\alpha\rangle \right\}
\end{equation}
for the quadrature $X_2$, where again the ``no prior information'' estimate, 
$\alpha_2 = {\rm Im}\,\alpha$, provides only half the input.  

Further insight into these best possible estimates is gained by expressing them 
solely in terms of the Husimi Q-function $Q(\alpha)$ in Eq.~(\ref{q}).  In 
particular, noting that variation with respect to $\alpha$ gives
\begin{eqnarray}
\delta \langle\alpha|\rho|\alpha\rangle & = & \langle\alpha| D(\delta\alpha 
)^\dagger \rho D(\delta\alpha)|\alpha\rangle - \langle\alpha|\rho|\alpha \rangle 
\nonumber \\ \label{vary}
& = & \langle\alpha|[\rho,a^\dagger]|\alpha\rangle\,\delta\alpha +
\langle\alpha|[a,\rho]|\alpha\rangle\,\delta\alpha^* ,
\end{eqnarray}
one may replace $a\rho$ by $[a,\rho]+\rho a$ in Eqs.~(\ref{bestcoh1}) and 
(\ref{bestcoh2}) to obtain
\begin{equation} \label{qest}
\tilde{X}_{j,{\rm opt}}(\alpha|\rho) = \alpha_j + 
(1/4)(\partial/\partial\alpha_j) \log Q 
\end{equation}
for $j=1,2$.  Hence  the best possible estimates differ significantly from 
$\alpha_1$ and $\alpha_2$ precisely when the gradient of the logarithm of the 
probability distribution, at the point corresponding to the measurement outcome, 
is large.

As examples, consider the cases where the field is known to be in a coherent 
state $|\beta\rangle$, and in a number state $|n\rangle$.  One then finds from 
Eqs.~(\ref{bestcoh1}) and (\ref{bestcoh2}), or equivalently from 
Eq.~(\ref{qest}),
\begin{eqnarray*}
\tilde{X}_{1, {\rm opt}}(\alpha|\beta) + i\tilde{X}_{2, {\rm opt}}(\alpha|\beta) & = & 
\frac{1}{2} (\alpha + \beta) ,\\
\tilde{X}_{1, {\rm opt}}(\alpha|n) + i\tilde{X}_{2, {\rm opt}}(\alpha|n) & = & \frac{1}{2} 
\,\alpha\, (1+ n/|\alpha|^2) 
\end{eqnarray*}
respectively.

Finally, to preview the effect of prior information on joint-measurement 
uncertainty relations, the uncertainties of the estimates ${\cal X}_{1,{\rm 
opt}}$ and ${\cal X}_{2,{\rm opt}}$ will be calculated here for the above 
coherent-state example.  These estimates are equivalent to the measurement of 
$({X}_{1,J}+\beta)/2$ and $({X}_{2,J}+\beta)/2$ respectively, and hence, using 
Eq.~(\ref{jointquad}),
\[ {\rm Var} {\cal X}_{1,{\rm opt}} = (1/4)\, {\rm Var}{X}_{1,J}=({\rm Var} X_1 + 
{\rm Var}Y_1)/4 = 1/8 , \]
with a similar result for ${\rm Var} {\cal X}_{2,{\rm opt}}$.  One therefore 
obtains the uncertainty product
\begin{equation} \label{uncoh}
\Delta {\cal X}_{1,{\rm opt}} \,\Delta {\cal X}_{2,{\rm opt}} = 1/8 
\end{equation}
for this example, which is {\it four times} better than the corresponding 
product,
\[ \Delta {X}_{1,J} \Delta {X}_{2,J} = 1/2 , \]
for the case when no prior information about the state is available.
It will be shown in the following section that this factor of 4 improvement is 
the ultimate limit.

\section{Uncertainty relations for optimal estimates}

\subsection{Dispersion vs inaccuracy: a geometric uncertainty relation}

There are two types of contribution to the ``uncertainty'' of 
an estimate.  The first, {\it dispersion}, is related to the statistics of the 
estimate itself, whereas the second, {\it inaccuracy}, is related to how well 
the estimate does its job of estimating a given observable.  These two
types 
of uncertainty are to some degree complementary, and it will be seen that for 
optimal estimates they are linked by a very simple uncertainty relation.

To characterise dispersion, let ${\cal A}_f$ denote the observable 
corresponding to a general estimate of $A$ from 
measurement ${\cal M}$, where outcome $m$ of ${\cal M}$ corresponds to outcome 
$f(m)$ of ${\cal A}_f$.  The statistics of the estimate are completely determined 
by the statistics of ${\cal M}$ and the choice of $f$, and in particular 
the root mean square deviation of ${\cal A}_f$ may be calculated 
in the usual way as
\begin{equation} \label{rms}
(\Delta {\cal A}_f)^2 = \sum_ m p(m|\rho)\,f(m)^2 -\left[ \sum_m p(m|\rho)
\,f(m)\right]^2  , 
\end{equation}
where the outcome probability $p(m|\rho)$ is given by Eq.~({\ref{prob}).
This quantity will be used as a measure of the dispersion of the estimate.

To characterise the inaccuracy of the estimate ${\cal A}_f$, one requires a 
measure $\epsilon({\cal A}_f)$ of the degree to which the estimate differs from 
the observable being estimated.  In particular, it should be
nonnegative, and vanish in the case of 
a perfect estimate (i.e., ${\cal A}_f\equiv{\cal A}$). 
The statistical deviation used in Eq.~(\ref{natdev}) satisfies
these properties, and hence the quantity
\begin{equation} \label{epsilon} 
\epsilon({\cal A}_f) := D_\rho (A,{\cal A}_f) 
\end{equation} 
will be used as a measure of inaccuracy of the estimate.  
Note from Eq.~(\ref{noisy}) that, for Hermitian observables, this measure is 
just the mean deviation of the noise operator associated with the estimate 
\cite{appleby,ozawapla,ag}.  Note also that the optimal 
estimates of Sec.~II based on prior information about the state of the system 
are precisely those estimates having the minimum possible inaccuracy: 
$\epsilon({\cal A}_f)\geq 
\epsilon({\cal A}_{\rm opt})$.
 
It follows immediately from Eq.~(\ref{vardiff}) of the Appendix that 
\begin{equation} \label{varsum}
(\Delta A)^2 = (\Delta {\cal A}_{\rm opt})^2 + \epsilon({\cal A}_{\rm opt})^2 ,
\end{equation} 
i.e., the dispersion and inaccuracy of the best possible estimate form 
the sides of a right-angled triangle having hypotenuse $\Delta A$. 
Thus {\it there is a fundamental tradeoff between dispersion and 
inaccuracy, valid for any measurement} ${\cal M}$.
This tradeoff may be geometrically represented by the 
constraint that ${\cal A}_{\rm opt}$ lies on a circle (or hypersphere)
having diametrically opposed ``poles'' $A$ and $\langle A\rangle$.  
These poles correspond to the optimal estimates for 
${\cal M}\equiv {\cal A}$ (i.e., a perfect estimate) and ${\cal M}\equiv {\bf 1}$ (i.e., a trivial estimate)
respectively.  Alternatively, one may represent the tradeoff by a circle of radius $\Delta A$ in the
dispersion-inaccuracy plane, with zero inaccuracy and zero dispersion 
corresponding to the cases ${\cal M}\equiv {\cal A}$ and ${\cal M}\equiv
{\bf 1}$ respectively.

The above geometric property, and the standard 
uncertainty relation $\Delta A\,\Delta 
B \geq |\langle [A,B]\rangle|/2$, 
allows one to immediately write down a general 
uncertainty relation for the best possible estimates of two Hermitian operators 
$A$ and $B$ from an arbitrary POM measurement ${\cal M}$:
\begin{equation} \label{geom}
\left[(\Delta {\cal A}_{\rm opt})^2 + \epsilon({\cal A}_{\rm opt})^2 
\right]^{1/2} \left[ (\Delta {\cal B}_{\rm opt})^2 + \epsilon({\cal B}_{\rm 
opt})^2\right]^{1/2} \geq |\langle [A,B]\rangle |/2 .
\end{equation}
Thus, for a non-zero lower bound, {\it 
one cannot make both estimates arbitrarily accurate while
making the corresponding dispersions 
arbitrarily small, no matter what measurement
scheme is adopted}.  Note that the lower bound is 
achieved if and only if the system is in a minimum-uncertainty state of $A$ and 
$B$.

\subsection{Incompatibility implies inaccuracy}

One has the useful lower bound
\begin{equation} \label{accbound}
\epsilon({\cal A}_{\rm opt})^2 \geq \sum_m \frac{|{\rm tr}[\rho (AM_m - 
M_mA)]|^2}{4\,{\rm tr}[\rho M_m]}
\end{equation}
for the inaccuracy of the best possible estimate. Equality holds in the case that 
$\rho$ is pure and ${\cal M}$ is complete (i.e., with $M_m = |m\rangle\langle 
m|$ for all $m$), and hence in particular for the case of heterodyne detection 
with pure signal and imageband fields.  
Note that since the optimal estimate of $A$ has, by
definition, the best possible accuracy, the righthand side of
Eq.~(\ref{accbound}) in fact provides a lower bound for the inaccuracy of
{\it
any} estimate of $A$ from ${\cal M}$, and hence is universal.

The lower bound is non-trivial whenever $\langle  
[A,M_m]\rangle$ does not vanish for some $m$, i.e., whenever $A$ and
${\cal M}$ are             
incompatible for state $\rho$. Hence, {\it one can never make
a perfect estimate of one observable from the measurement
of a second incompatible observable}.   
When ${\cal A}$ and ${\cal 
M}$ are a pair of canonically conjugate observables, the lower bound is 
proportional to the Fisher information of ${\cal M}$,
and the case of equality corresponds to an ``exact uncertainty relation''  
for ${\cal A}$ and ${\cal M}$ \cite{hallfish, halleur}.

Eq.~(\ref{accbound}) generalises Eq.~(47) of Ref.~\cite{halleur} (in the
context of exact uncertainty relations) and Eq.~(14) in
Ref.~\cite{lars1} (in the context of weak values), to general POM
measurements
${\cal M}$. It follows via 
the Schwarz inequality $|{\rm tr}[K^\dagger L]|^2 \leq {\rm 
tr}[K^\dagger K] {\rm tr}[L^\dagger L]$, which gives 
\[ |{\rm tr}[\rho AM_m]|^2 \leq {\rm tr}[\rho M_m] {\rm tr}[\rho AM_mA] \]
for the choice $K=\rho^{1/2}M_m^{1/2}$, $L=\rho^{1/2}AM_m^{1/2}$.  Noting the 
first equality in Eq.~(\ref{vardiff}) of the Appendix, and using $(z+z^*)^2=
4|z|^2-|z-z^*|^2$ for 
$z = {\rm tr}[\rho AM_m]$ appearing in the optimal estimate in 
Eq.~(\ref{best}), then leads directly to Eq.~(\ref{accbound}).  Equality holds 
for $K$ proportional to $L$, and hence in particular for a complete measurement 
on a pure state. 

\subsection{Example: heterodyne detection}

For heterodyne detection with a vacuum-state imageband field, as discussed in 
Sec.~II.E, it will be shown that one has the further independent inequalities
\begin{eqnarray} \label{unbest}
\Delta {\cal X}_{1, {\rm opt}}\, \Delta {\cal X}_{2, {\rm opt}} & \geq & 1/8 ,\\ 
\label{accbest}
\epsilon({\cal X}_{1, {\rm opt}})^2 + \epsilon({\cal X}_{
2, {\rm opt}})^2 & \geq 
& 1/4 ,
\end{eqnarray}
for the dispersions and the inaccuracies of the best possible estimates.  The 
first relation is saturated for coherent states, and the second relation is 
saturated for all pure states.

Note that for the analogous case of a canonical joint measurement of position 
and momentum as discussed in Sec.~II.D (with the auxilary system in a minimum 
uncertainty state), it follows immediately from Eq.~(\ref{unbest}) that one has 
the corresponding uncertainty relation
\begin{equation} \label{uncanon}
\Delta {\cal X}_{{\rm opt}}\,\Delta {\cal P}_{{\rm opt}}\geq\hbar/4 ,
\end{equation}
improving on the``universally unbiased'' lower bound in Eq.~(\ref{un4}) by a 
factor of 4.  Thus, even when one has {\it complete} information about the system 
prior to measurement, there is still a fundamental lower bound to the product of 
the dispersions of the optimal estimates.

To prove Eqs.~(\ref{unbest}) and (\ref{accbest}), recall that the $2\times 2$ 
covariance matrix $C$ for two random variables ${A}_1$, ${A}_2$ is given by 
$C_{jk}:=\langle {A}_j{A}_k\rangle -\langle {A}_j \rangle\langle {A}_k\rangle$.  
Hence 
the covariance matrix $C^{\rm opt}$ of the optimal quadrature estimates follows 
via Eqs.~(\ref{x0}), (\ref{p0}) and (\ref{qest}) as
\begin{eqnarray}
C^{\rm opt}_{jk} & = & \langle \alpha_j\alpha_k\rangle + \frac{1}{4} \int 
d^2\alpha \, \left(\alpha_j\frac{\partial Q}{\partial \alpha_k} + 
\alpha_k\frac{\partial Q}{\partial \alpha_j} \right) + 
\frac{1}{16}F^Q_{jk} \nonumber 
\\ \label{mat}
& = & C^{Q}_{jk} + (1/16)F^Q_{jk} - (1/2)\delta_{jk} . 
\end{eqnarray}
Here $C^{Q}$ is the covariance matrix for the joint-quadrature observables 
$X_{1,J}$ and $X_{2,J}$ in Eq.~(\ref{jointquad}), $F^Q$ denotes the Fisher 
information matrix of the Husimi Q-function with 
\cite{cover} 
\begin{equation} \label{fish}
F^Q_{jk} := \int d^2\alpha\, Q\,(\partial \log Q/\partial\alpha_j) (\partial 
\log Q/\partial\alpha_k) , 
\end{equation}
and integration by parts has been used to obtain the second line.

Now, if $F_j$ denotes the Fisher information of the marginal distribution 
$Q_j(\alpha_j)$ for $\alpha_j$, then the Cramer-Rao inequality from classical 
statistics \cite{cover} yields $F_j \geq 1/C^Q_{jj}$.  One also has
\[
0 \leq \int d^2\alpha\, Q(\alpha)\,[(\partial \log Q/\partial \alpha_j) -  
(\partial \log Q_j/\partial \alpha_j)]^2 = F^Q_{jj} - F_j . \]
Substitution of these inequalities into Eq.~(\ref{mat}) then yields
\[ C^{\rm opt}_{jj} \geq C^{Q}_{jj} + 1/(16C^{Q}_{jj}) -1/2. \]
Writing ${\rm Var} X_1=\gamma r/4$, ${\rm Var} X_2=\gamma /(4r)$, with 
$\gamma\geq 1$ (to satisfy the standard uncertainty relation for the 
quadratures) and $r\geq 0$, and noting from Eq.~(\ref{jointquad}) that 
$C^{Q}_{jj} = {\rm Var} X_j + 1/4$, therefore leads to
\[ C^{\rm opt}_{11} C^{\rm opt}_{22} \geq \frac{\gamma^3}{16(\gamma^2
-1)} \left[\frac{\gamma}{\gamma +r} - \frac{1}{\gamma r+1} \right] . \]
Minimising the righthand side with respect to $r$ gives $r=1$; minimising the 
resulting expression with respect to $\gamma\geq1$ then gives $\gamma=1$; and 
Eq.~(\ref{unbest}) immediately follows.

Finally, to obtain Eq.~(\ref{accbest}), note first that combining 
Eqs.~(\ref{q}), (\ref{vary}), (\ref{accbound}) and (\ref{fish}) gives 
\begin{equation} \label{fishbound}
\epsilon({\cal X}_{1,{\rm opt}})^2 \geq F^Q_{22}/16, \,\,
\,\,\,\,\,\,\,\,\,\,\,\, \epsilon({\cal X}_{2,{\rm opt}})^2 \geq F^Q_{11}/16
,
\end{equation}
where equality holds for all pure states.  Thus the accuracy of the estimate of 
one quadrature is related to the Fisher information of the other quadrature.  
Moreover, taking the trace of $C^{\rm opt}$ in Eq.~(\ref{mat}) and using the 
Euclidean relation in Eq.~(\ref{varsum}), one also has
\[ \epsilon({\cal X}_{1,{\rm opt}})^2 + \epsilon({\cal X}_{2,{\rm opt}})^2 = 1/2 
- (F^Q_{11} + F^Q_{22})/16   \]
(giving an {\it upper} bound of half a photon for the lefthand side).
Comparison with Eq.~(\ref{fishbound}) immediately yields the known relation 
\cite{cover}
\begin{equation} \label{tracefish}
F^Q_{11} + F^Q_{22} \leq 4
\end{equation}
for the trace of the Fisher information matrix, which when inserted back into 
the previous expression yields Eq.~(\ref{accbest}) as desired.

\section{Universal joint-measurement uncertainty relation} 

\subsection{Arbitrary estimates}

The uncertainty relation to be derived in this section applies to {\it any} 
estimates of two Hermitian operators $A$ and $B$ from a general measurement 
${\cal M}$.  Unlike the geometric uncertainty relation of the previous section, 
it is valid for both optimal and non-optimal estimates, and is independent of 
whether or not any prior information about the system is available.  
The associated derivation may be modified to obtain the more restrictive 
uncertainty relations satisfied by universally unbiased estimates, such as 
Eq.~(\ref{un4}).

Suppose then that $f(m)$ and $g(m)$ are general estimates for $A$ and $B$ 
respectively, for measurement result ${\cal M}=m$. These estimates thus 
correspond to two compatible observables ${\cal A}_f$ and ${\cal B}_g$, measured 
by measuring ${\cal M}$ and for outcome $m$ assigning the values $f(m)$ and 
$g(m)$ respectively.  It will be shown that these estimates satisfy the 
universal uncertainty relation 
\begin{equation} \label{ungen}
\Delta {\cal A}_f\,\epsilon({\cal B}_g) + \epsilon({\cal A}_f)\,
\Delta {\cal B}_g 
+  \epsilon({\cal A}_f)\, \epsilon({\cal B}_g) \geq |\langle [A,B]\rangle| 
/2 .
\end{equation}
This uncertainty relation is therefore a fundamental expression of the 
limitations imposed by complementarity on quantum systems.

As a very simple example, suppose that one makes no physical measurement at
all, but simply estimates $A=0$ and $B=0$ on every occasion. Then
clearly the dispersions of the estimates vanish: $\Delta {\cal
A}_f=\Delta {\cal B}_g$=0.  The universal uncertainty
relation Eq.~(\ref{ungen}) then implies that the product of the
inaccuracies of such trivial estimates is {\it non-trivially} 
bounded below, i.e.,
\[ \epsilon({\cal A}_f)\,\epsilon({\cal B}_g) \geq |\langle
[A,B]\rangle|
/2 .\]

As a less trivial example, suppose that the position $X$ of a 
quantum particle is 
measured, and used to estimate both the position and the momentum of the 
particle.  It is natural to choose ${\cal X}_f \equiv  X$ (this is in fact the 
{\it optimal} estimate, whether or not any prior information is available).  
This estimate of $X$ is perfectly accurate, i.e., $\epsilon( {\cal X}_{\rm f}
)=0$, and 
hence from Eq.~(\ref{ungen})
 \[ \Delta X\, \epsilon({\cal P}_g)\geq \hbar/2 \]
for {\it any} corresponding estimate ${\cal P}_g$ of the momentum.  
Note that this is a 
stronger result than the corresponding geometric uncertainty relation following 
from Eq.~(\ref{geom}).

The proof of Eq.~(\ref{ungen}) proceeds via a formal trick - the representation 
of the measurement ${\cal M}$ as a Hermitian operator $M^\prime$ on an extended 
Hilbert space.  This representation (a Naimark extension) preserves the 
statistical deviation between observables, while allowing one to 
exploit algebraic properties of Hermitian operators.  Any such representation 
can be used for the proof, however, the choice of a product space representation 
is perhaps the simplest.

In particular, for a given POM ${\cal M}\equiv\{M_m\}$ one can always (formally) 
introduce an auxilary system described by some fixed state $\rho'$, and a 
Hermitian operator $M^\prime$ acting on the tensor product of the system and 
auxilary system Hilbert spaces, such that the statistics of ${\cal M}$ and 
$M^\prime$ are identical \cite{helstrom,holevo, lahti}, i.e.,
\begin{equation} \label{rep}
p(m|\rho) = {\rm tr}[\rho M_m] = {\rm tr}[\rho \otimes\rho' 
M^\prime_m] 
\end{equation}
for all density operators $\rho$ and outcomes $m$, where $M^\prime_m$ denotes 
the projection on the eigenspace associated with eigenvalue $m$ of $M^\prime$.  
Note that this representation is used here as a formal mathematical device only, 
with no physical content.  

It follows immediately from Eq.~(\ref{rep}) that the statistics of general 
estimates ${\cal A}_f$ and ${\cal B}_g$ are equivalent to the statistics of the 
(commuting) Hermitian operators $f(M^\prime)$ and $g(M^\prime)$ respectively.  
Further, if $\{|s'\rangle\}$ denotes a complete set of kets for the auxilary 
Hilbert space, Eq.~(\ref{rep}) yields the partial trace relation 
\begin{equation} \label{partial}
{\rm tr}_{\rho'}[\rho'M^\prime_m] := \sum_{s'} \langle s'|\rho' 
M^\prime_m|s'\rangle = M_m . 
\end{equation}
Hence, using Eqs.~({\ref{dev}) and (\ref{gendev}) of the Appendix, one has
\begin{eqnarray*}
D_{\rho\otimes\rho'}(A,f(M^\prime))^2 & = & \langle A^2\rangle + \langle 
f(M^\prime)^2\rangle - \sum_m f(m) {\rm tr}[\rho\otimes\rho' (AM^\prime_m + 
M^\prime_mA)]\\
& = & \langle A^2\rangle + \langle {\cal A}_f^2\rangle - 
\sum_m f(m) \left\{{\rm 
tr}_\rho[\rho A\, {\rm tr}_{\rho'}[\rho'M^\prime_m]] + c.c. \right\}\\
& = & \langle A^2\rangle + \langle {\cal A}_f^2\rangle - {\rm tr}[\rho 
A\overline{A}_f + \overline{A}_fA\rho]\\
& = & D_\rho(A, {\cal A}_f)^2 = \epsilon({\cal A}_f)^2
\end{eqnarray*}
and thus the representation preserves statistical deviation and inaccuracy. 
Writing $\delta A=A-f(M^\prime)$, $\delta B = B - g(M^\prime)$, it follows 
that $\epsilon({\cal A}_f)^2 = \langle(\delta A)^2\rangle$ and 
$\epsilon({\cal B}_g)^2 = \langle(\delta B)^2\rangle$, and hence that
\begin{eqnarray*}
|\langle [A,B]\rangle| & = & |\langle [f(M^\prime)+\delta A,g(M^\prime) + 
\delta B]\rangle|\\
& \leq & |\langle [f(M^\prime),\delta B]\rangle| + |\langle [\delta A, 
g(M^\prime)]\rangle| + |\langle [\delta A,\delta B]\rangle|\\
& \leq & 2\Delta f(M^\prime)\epsilon({\cal B}_g)
+ 2\epsilon({\cal A}_f) \Delta g(M^\prime) + 2\epsilon({\cal A}_f)
 \epsilon({\cal B}_g) ,
\end{eqnarray*}
using the triangle inequality and the Schwarz inequality $\langle (K-k)^2\rangle 
\langle L^2\rangle \geq |\langle[K,L]\rangle|^2/4$ (in a manner formally similar 
to Ozawa's proof of Eq.~(\ref{un3}) 
\cite{ozawapla,ozawa}).  The last line is equivalent to the universal 
uncertainty relation in Eq.~(\ref{ungen}).  

Finally, the above derivation may be modified to obtain a stronger uncertainty 
relation, valid for the special case of {\it universally unbiased} estimates  of 
$A$ and $B$  \cite{appleby,ak, ag}.
In particular, the requirements that $\langle{\cal A}_f \rangle = \langle 
A\rangle$, $\langle{\cal B}_g \rangle = \langle B\rangle$ for {\it all} states 
$\rho$ implies via Eq.~(\ref{partial}) that 
\[A = {\rm tr}_{\rho'}[\rho' f(M^\prime)],\,\,\,\,\,\,
B = {\rm tr}_{\rho'}[\rho' g(M^\prime)] . \] 
Hence ${\rm tr}_{\rho'}[\rho'Ag(M^\prime)] = AB = {\rm tr}_{\rho'}[\rho' 
f(M^\prime)B)]$, implying that $\langle [\delta A,\delta B]\rangle = \langle -
[A,B]\rangle$.  Thus, with no triangle inequality being necessary, the Schwarz 
inequality yields
\begin{equation} \label{uni}
\epsilon({\cal A}_f)\,\epsilon({\cal B}_g) \geq |\langle [\delta A, \delta 
B]\rangle|/2 = |\langle [A,B]\rangle|/2 .
\end{equation}
The joint uncertainty relation for universally 
unbiased joint measurements of position and 
momentum, Eq.~(\ref{un4}), is a straightforward consequence of this result 
\cite{appleby,ak, ag}.

\subsection{Example: EPR estimates}

The notion that the properties of position and motion are incompatible goes back 
nearly 2500 years to Zeno of Elea (who resolved the issue by concluding that 
motion was impossible).  However, in classical physics this notion was rejected 
due to the existence of a consistent model: one can simultaneously define both 
the position and motion of a classical system by assuming that it follows a 
(differentiable) continuous trajectory in configuration space.  Unfortunately, 
in the standard quantum formalism there are no such trajectories for physical 
systems, and a new resolution of the issue is needed.  

In the standard interpretation of quantum mechanics, 
as formulated by Heisenberg 
and Bohr \cite{heisp20, bohr}, one takes the view that the properties of 
position and motion are indeed incompatible, in the sense of being unable to be 
accurately defined/measured simultaneously, and to this extent agrees with Zeno.
However, others (most notably Einstein) have argued that the quantum formalism 
is in fact {\it incomplete}, 
and that quantum systems can in particular have simultaneously 
well-defined physical values of position and momentum \cite{epr}.  It has since 
been shown that any such ``hidden variable'' interpretation requires the 
existence of a mutual influence or 
conspiracy  between a measurement made on one 
system and the values ascribed to a space-like separated system 
\cite{bell,shim,esp}.  Even so, it is of interest to consider the relation of 
the famous incompleteness argument made by Einstein, Podolsky and Rosen (EPR) 
\cite{epr} to the principle of complementarity, as embodied in 
Eq.~(\ref{ungen}).

The EPR paper considers two particles described by an eigenket of relative 
position and total momentum \cite{epr}.  Clearly, the position of the first 
particle can be estimated precisely by a direct measurement of the position, 
with perfect accuracy: 
$\epsilon({\cal X}_{\rm opt}) = 0$.  Simultaneously, the  
correlation between the particles allows the momentum of the first particle to 
also be estimated precisely, from a measurement of the momentum of the second 
particle, again with perfect accuracy: $\epsilon({\cal P}_{\rm opt}) = 0$.  At 
first sight it thus appears that the universal joint-measurement uncertainty 
relation in Eq.~(\ref{ungen}) is violated by the EPR example.  

To see what is happening, it is helpful to replace the non-normalisable eigenket 
considered by EPR with the physical wavefunction 
\[ \psi(x,x') = K e^{-(x-x'-a)^2/4\sigma^2} e^{-\tau^2(x+x')^2/4\hbar^2} 
e^{ip_0(x+x')/2\hbar} ,\]
where $K$ is a normalisation constant and $\sigma,\tau \ll 1$ in suitable units.  
One has
\[ \langle X-X'\rangle = a,\,\,\,\,\,\,\,\,\,\,\, {\rm Var}(X-X') = \sigma^2 \ll 
1 , \]
\[ \langle P+P'\rangle = p_0,\,\,\,\,\,\,\,\,\,\,\, {\rm Var}(P+P') = \tau^2 \ll 
1 , \]
and hence $\psi$ is an {\it approximate} eigenstate of the relative position and 
total momentum.

Suppose now that $X$ and $P'$ are simultaneously measured as before, with 
measurement results $x$ and $p'$ respectively.  The corresponding best possible 
estimates of $X$ and $P$ then follow via Eq.~(\ref{bestpure}) as
\[ \tilde{X}_{\rm opt} = x, \,\,\,\,\,\,\, 
\tilde{P}_{\rm opt} = \frac{\hbar^2(p_0-p') + 
\sigma^2\tau^2p'}{\hbar^2 + \sigma^2\tau^2}\approx p_0-p' . \]
The dispersions and inaccuracies of these estimates follow from straightforward 
calculation as
\[ \Delta {\cal X}_{\rm opt} = (\hbar^2+\sigma^2\tau^2)^{1/2}/(2\tau)  \approx 
\hbar/(2\tau),\,\,\,\,\,\,\,\epsilon({\cal X}_{\rm opt}) = 0, \]
\[ \Delta {\cal P}_{\rm opt} = \frac{|\hbar^2 - \sigma^2\tau^2|}{ 2\sigma 
(\hbar^2 + \sigma^2\tau^2)^{1/2}}  \approx \hbar/(2\sigma),\,\,\,\,\,\, 
\epsilon({\cal P}_{\rm opt}) = \frac{\hbar\tau}{ (\hbar^2 + 
\sigma^2\tau^2)^{1/2}} \approx \tau . \]
Substitution into the lefthand side of the joint measurement uncertainty 
relation in Eq.~(\ref{ungen}) then gives 
$\hbar/2$, which is precisely equal to the value of 
the righthand side - the state is in fact a minimum joint-uncertainty
state of position and momentum (other equalities for this state are
given in Ref.~\cite{halleur}, where the effect of wavefunction collapse
on optimal estimates is also considered).  

The above results support, in a quantitative manner, Bohr's defence of the 
consistency of complementarity with the completeness of the standard quantum 
formalism \cite{bohr, bohrepr}. The EPR argument in fact goes somewhat further, 
asserting the physical reality of the estimated value of $P$ from the 
measurement of $P'$, and the simultaneous physical reality of the estimated 
value of $X$ following from the alternative measurement of $X'$ \cite{epr}.  
However, precisely because these measurements do not refer to a single 
experimental setup, such assertions go beyond the quantum formalism, and cannot 
be tested via Eq.~(\ref{ungen}).

More generally, even 
when one has full knowledge of the state of some  system, and 
uses this prior information to make the best possible estimates of two 
complementary observables from a given experimental setup, there remains a 
fundamental tradeoff between dispersion and inaccuracy - embodied by the 
universal uncertainty relation in Eq.~(\ref{ungen}) - which prevents  
simultaneous knowledge of the corresponding physical properties.

\subsection{Example: linear estimates}

It is of interest to consider an example where one does not know the state of 
the system before measurement, but does have prior knowledge of the averages of 
one or more observables.  While such prior information is by itself insufficient to make 
an optimal estimate as per Eq.~(\ref{best}), it can still be taken into account 
to improve on the ``no information'' estimate of Eq.~(\ref{bestnoinf}). 
 
One method of proceeding might be to introduce some physical principle to 
assign a unique state to the system that is consistent with the given prior 
information, and to calculate estimates by substituting this state for $\rho$ 
in Eq.~(\ref{best}).  For example, the maximum entropy principle of Jaynes 
could be used for this purpose 
\cite{jaynes} (indeed the ``thermodynamic'' example in Eq.~(\ref{thermal}) 
may be reinterpreted in this way, where the form of the density operator 
corresponds to the maximum entropy state consistent 
with a known prior average energy of the system \cite{jaynes}).  

In general, however, there are many possible physical states consistent with given prior knowledge about certain averages.  Further, the available prior information may well imply, for example, that the system is {\it not} described by a maximum entropy state (eg, in a communication setup it may be known that each signal is described by one of a number of fixed pure states $|\psi_1\rangle, |\psi_2\rangle,\dots$ having equal average energies). It is therefore important to consider estimation methods that use only the prior information that is available, without requiring assumptions about the actual state of the system.  Here {\it linear} estimates and their joint uncertainty properties will be examined.

Consider first a detection system for a classical signal $s$, which is subject 
to uncorrelated noise $n$, resulting in a measured signal $m=s+n$.  It will be 
assumed that $\langle n\rangle =0$. If $m$ is taken as an estimate for $s$, the 
average deviation of this estimate from the actual signal $s$ is quantified by
\[ \epsilon^2 = \langle \,(m-s)^2\,\rangle = N , \]
where $N$ denotes the noise variance $\langle n^2\rangle$.  

However, one can do better if there is some prior information about the signal 
statistics.  For example, suppose one knows the average 
value $\overline{s}=\langle s\rangle$ and the variance $S=\langle (s-
\overline{s})^2\rangle$ of the signal.  Then it is straightforward to show that 
the {\it linear} estimate $m_{\rm lin}=\lambda m +(1-\lambda)\overline{s}$ has a 
minimum statistical deviation from the signal $s$ given by
\[ \epsilon_{\rm lin}^2= \langle \,(m_{\rm lin} -s)^2\,\rangle = NS/(S+N) < 
\epsilon^2 , \]
corresponding to the choice $\lambda = S/(S+N)$.  The associated
rms uncertainty of this estimate follows as 
\[ \Delta m_{\rm lin}= S/(S+N)^{1/2} = (1+N/S)^{-1}\Delta m. \]
Thus, use of the prior information allows not only a {\it better} estimate of 
the signal, but also a reduction in the {\it dispersion} of the estimate of the 
signal.  Note that for the particular case of {\it Gaussian} signal and
noise distributions, the above linear estimate is in fact optimal over 
any other estimate \cite{hancock}. 

Consider now the canonical joint measurement of position and momentum for a 
quantum particle as previously discussed, corresponding to measurement of the 
commuting operators $X_J=X+X'$, $P_J=P-P'$, where the primed variables refer to 
an auxilary particle in a minimum uncertainty state with $\langle X'\rangle 
=\langle P'\rangle = 0$.  
It will be assumed that all that is known about the particle prior to 
measurement are the means and variances of $X$ and $P$.

The observables $X$, $X'$, $X_J=X+X'$ all commute, and are therefore completely 
analogous to the respective classical variables $s$, $n$, and $m=s+n$.  It 
immediately follows that the best linear estimate of $X$ from $X_J$, given 
knowledge of $\langle X\rangle$ and ${\rm Var} X$, is equivalent to measurement 
of the operator $X_{\rm lin}=\lambda X_J + (1-\lambda)\langle X\rangle$, with 
$\lambda = (1+{\rm Var} X'/{\rm Var} X)^{-1}$;
associated inaccuracy
\[ \epsilon(X_{\rm lin}) = \Delta X\,\Delta X'/( {\rm Var} X + {\rm
Var} X')^{1/2} ; \]
and associated dispersion
\[ \Delta X_{\rm lin} = {\rm Var} X/({\rm Var} X + {\rm Var} X')^{1/2} =  
(1+{\rm Var} X'/{\rm Var} X)^{-1}\Delta X_J . \]
One similarly has an optimal linear estimate $P_{\rm lin}$ obtained from  
knowledge of $\langle P\rangle$ and ${\rm Var} 
P$, with analogous expressions for $\epsilon(P_{\rm
lin})$ and $\Delta P_{\rm lin}$.  

Note that there is a degree of freedom remaining, which may be tuned
for further optimality.  In particular, the squeezing ratio $\Delta X'/\Delta
P'$ may be chosen to minimise some suitable cost function.  For example,
for a harmonic oscillator 
one might choose to minimise the ``inaccuracy
energy'' $\epsilon(P_{\rm lin})^2/(2m) + (m\omega^2/2)\epsilon(X_{\rm lin})^2$. 
However, the existence of the universal uncertainty
relation in Eq.~(\ref{ungen}) suggests the more generic ``joint
uncertainty'' cost function
\[ J = \Delta X_{\rm lin} \,\epsilon(P_{\rm lin}) + \epsilon(X_{\rm
lin})\, \Delta P_{\rm lin} + \epsilon(X_{\rm lin})\,\epsilon(P_{\rm lin})
.\]

Minimising $J$ with respect to the squeezing ratio leads to two regimes.
First, if $\Delta X\,\Delta P \leq 2\hbar$, then it is optimal to choose
$\Delta X'/\Delta P' = \Delta X/\Delta P$, which leads to the
inequality 
\[ \Delta X_{\rm lin}\,\Delta P_{\rm lin} \geq [1+\hbar^2/(4{\rm Var}
X\,{\rm Var} P)]^{-1}\,\hbar/2\geq\hbar/4, \] 
analogous to the lower bound in Eq.~(\ref{uncanon}).
However, for $\Delta X\,\Delta P>2\hbar$, it is optimal to choose either
of $\Delta X'$ and $\Delta P'$ equal to zero, corresponding to the
alternatives  
$X_{\rm lin}=X$, $P_{\rm lin}=\langle P\rangle$ and $X_{\rm
lin}=\langle X\rangle$, $P_{\rm lin}=P$ respectively 
- i.e., not to bother with a true joint measurement at all! 
A similar dichotomy of regimes has been noted previously for the special case of
Gaussian states \cite{holevo2}.

\section{Conclusions}

A general formula for the best possible estimate of one observable from the 
measurement of another has been given, and applied in a number of
settings.  A universal joint-measurement uncertainty relation has also
been given, which
quantifies the principle of 
quantum complementarity for all possible experimental setups.  
Describing 
measurements by  completely general POMs (which require only that 
probabilities are positive and sum to unity), implies that the main 
results of the paper are universally applicable, and independent of any 
dynamical models and interpretational issues concerning quantum measurement. It 
is also worth noting that the use of a general POM includes the case where
an experimenter bases an estimate 
on the results of a {\it plurality} of measurements, 
obtained by carrying out a number of (predetermined) consecutive physical 
operations (described by ``completely positive'' linear maps 
\cite{lahti}).

It has been shown that by using prior information about the system (eg, the 
state of the system in Sec.~III.C and the mean and variance of certain 
observables in Sec.~IV.C) one can improve the standard uncertainty relation for 
the canonical joint measurement of position 
and momentum by up to a factor of 4.
However, unlike the classical case, 
if one makes optimal use of {\it complete} information 
about the system before measurement, one cannot do any better than this 
- complementarity cannot be circumvented by the use of prior knowledge.  The 
principle of complementarity is similarly consistent with respect to the 
properties of entangled systems -  as demonstrated in Sec.~IV.B, quantum 
correlations cannot be exploited to violate the universal joint-measurement 
uncertainty relation of Eq.~(\ref{ungen}).  

Finally, it would be of interest to determine the best possible estimate of an 
observable under the imposition of further natural restrictions.  For example, 
one could require that an estimate of photon number, from some general 
measurement, minimise statistical deviation 
subject to the further constraint of 
being a positive integer.  This would reduce the accuracy of the estimate 
relative to the unconstrained case, 
but has the advantage of incorporating prior 
information about the possible physical values of the observable being 
estimated.  It would similarly be of interest to consider alternative
characterisations of dispersion and inaccuracy (eg, entropy and relative entropy).

Some time after this paper was submitted, a related eprint by Ozawa has appeared
\cite{ozawacon}, giving an independent derivation of the universal uncertainty
relation in Eq.~(\ref{ungen}).


\appendix

\section{}

The proofs of Eqs.~(\ref{best}) and (\ref{bestnoinf}), for optimal estimates of 
a Hermitian operator $A$ from a 
general measurement ${\cal M}$, are given here.  
The generalisation to the optimal estimate of {\it any} POM observable ${\cal 
A}$ from measurement of ${\cal M}$ is also discussed.

The main ingredient required is a measure of ``how good'' a given estimate of 
$A$ is.  For the case of two Hermitian operators $A$ and $B$, a natural measure of  
how well one mimics the other, for a given state $\rho$, is 
given by the statistical deviation
\begin{equation} \label{dev}
D_\rho (A,B)^2 = {\rm tr}[\rho (A-B)^2] .
\end{equation}
This measure was used in the proof of Eq.~(\ref{bestpure}) for the special case 
where ${\cal M}$ corresponds to a Hermitian operator $M$.  However, to consider 
arbitrary measurements ${\cal M}$ 
it is necessary to generalise this measure to 
the case where one observable is an arbitrary POM observable.

Fortunately, the generalisation of Eq.~(\ref{dev}) is quite straightforward 
\cite{ozawa,hallalg}.  In particular, it is natural to define the statistical 
deviation between a Hermitian operator $A$ and a POM observable ${\cal M} = 
\{M_m\}$ by
\begin{equation} \label{gendev}
D_\rho (A, {\cal M})^2 = \sum_m {\rm tr}[M_m(A-m)\rho (A-m)]
= {\rm tr}[\rho(A-\overline{M})^2] + {\rm tr}[\rho(\overline{M^2} - 
\overline{M}^2)] ,
\end{equation}
where $\overline{M^j} := \sum_m m^j\,M_m$.  This expression reduces to 
Eq.~(\ref{dev}) for Hermitian observables.  It follows directly from a natural 
algebra for POM observables \cite{hallalg} 
(being the square root of the average of the
square of the ``difference'' of two such obervables), 
and has also been postulated {\it ab 
initio} in Ref.~\cite{ozawa}.  It first 
appeared in the context of estimation of 
photon number from an optical phase measurement \cite{hallphaseeprint}.  

To obtain Eq.~(\ref{best}), let ${\cal A}_f$ denote the observable 
corresponding to a 
general estimate of $A$ from ${\cal M}$, where outcome $m$ of ${\cal M}$ 
corresponds to outcome $f(m)$ of ${\cal A}_f$.  The statistical deviation 
between ${\cal A}_f$ and $A$ follows from the first equality in 
Eq.~(\ref{gendev}) as 
\begin{eqnarray*}
D(A, {\cal A}_f)^2 & = & \langle\,A^2\,\rangle -\sum_m f(m)\, {\rm 
tr}[\rho(AM_m + M_mA)] + \sum_m f(m)^2 {\rm tr}[\rho M_m]\\
& = & \langle\,A^2\,\rangle -2\sum_m f(m)
\tilde{A}_{\rm opt}(m|\rho)\, {\rm tr}[\rho M_m] 
+ \sum_m f(m)^2 {\rm tr}[\rho M_m]\\
& = & \langle\,A^2\,\rangle - \sum_m \tilde{A}_{\rm opt}(m|\rho)^2\, {\rm tr}[\rho M_m] 
+ \sum_m [f(m)- \tilde{A}_{\rm opt}(m|\rho)]^2 {\rm tr}[\rho M_m],
\end{eqnarray*}
where $\tilde{A}_{\rm opt}(m|\rho)$ is the estimate defined in Eq.~(\ref{best}).  The 
last term is nonnegative, and hence the statistical deviation is minimised by 
the choice $f(m) = \tilde{A}_{\rm opt}(m|\rho)$, as per Eq.~(\ref{best}).  
Note that 
choosing ${\cal A}_f = {\cal A}_{\rm opt}$ in the above expression, and using 
Eq.~(\ref{av}), gives 
\begin{equation} \label{vardiff}
D_\rho(A, {\cal A}_{\rm opt})^2 = \langle\, A^2\,\rangle - \langle \,{\cal 
A}_{\rm opt}^2\,\rangle = {\rm Var} A - {\rm Var} {\cal A}_{\rm opt} 
= D_\rho(A, \langle A\rangle)^2 - D_\rho({\cal A}_{\rm opt}, \langle
A\rangle)^2
\end{equation}
for the minimum statistical deviation.

The proof of Eq.~(\ref{bestnoinf}) is completely analogous, where the 
statistical deviation in Eq.~(\ref{gendev}) is replaced by the generalised 
Hilbert-Schmidt distance 
\begin{eqnarray}
d(A,{\cal M})^2 & := & \sum_m {\rm tr}[M_m(A-m)^2] \nonumber\\ \label{dist}
& = & {\rm tr}[(A-\overline{M})^2] + {\rm tr}[(\overline{M^2} - \overline{M}^2)] 
, 
\end{eqnarray}
obtained via a natural algebra for POM observables \cite{hallalg}.
Note that this measure is proportional to the average of the square of
the statistical 
deviation over all states.

Finally, it may be asked whether one can define the best possible estimate when 
${\cal A}$ does {\it not} correspond to a Hermitian operator.  This is of 
interest, for example, if one wants to make the best estimate of elapsed time  
or optical phase from the measurement of some observable such as position or 
photon number.  It turns out that the generalisation of statistical deviation is 
highly non-trivial in this case, as certain consistency conditions must be 
satisfied \cite{hallalg}.  However, for the special case of {\it complete} 
observables  ${\cal A}$ and ${\cal M}$ (i.e., with $A_a = |a\rangle\langle a|$, 
$M_m = |m\rangle\langle m|$), which further satisfy the condition that no two 
kets from the combined set $\{ |a\rangle,|m\rangle\}$ are proportional, it 
follows from Sec.~4 of \cite{hallalg} that the statistical deviation has the 
simple generalised form
\[ D_\rho({\cal A}, {\cal M})^2 = 
{\rm tr}[\rho(\overline{A^2} + \overline{M^2} - 
\overline{A}\,\,\overline{M} - \overline{M}\,\,\overline{A} )] , \]
with $\overline{A^j}$ and $\overline{M^j}$ defined as above.  It may be shown 
that the best possible estimate of ${\cal A}$, from a measurement result $m$ of 
${\cal M}$ on a known state $\rho$, follows in this case as
\begin{equation} \label{bestgen}
\tilde{A}_{\rm opt}(m|\rho) = \frac{\langle m| \rho \overline{A} + \overline{A}\rho 
|m\rangle}{2\langle m|\rho |m\rangle} .
\end{equation}
However, more generally one cannot simply replace $A$ by $\overline{A}$ in 
Eq.~(\ref{best}).


\begin{thebibliography}{99}

\bibitem{helstrom} C.W. Helstrom, {\it Quantum Detection and Estimation Theory} 
(Academic Press, New York, 1976).
\bibitem{holevo} A.S. Holevo, {\it Probabilistic and Statistical Aspects of 
Quantum Theory} (North-Holland, Amsterdam, 1982).
\bibitem{caves} S.L. Braunstein, C.M. Caves, and G.J. Milburn, Ann. Phys. (N.Y.) 
{\bf 247}, 135 (1996).
\bibitem{heisp20} W. Heisenberg, {\it The Physical Principles of the Quantum 
Theory} (Dover, USA, 1930).  The quote is from page 20.
\bibitem{aharonov} Y. Aharonov and D. Bohm, Phys. Rev. {\bf 122}, 1649 (1961).
\bibitem{appleby}  D.M. Appleby, Int. J. Theor. Phys. {\bf 37}, 1491 (1998).
\bibitem{dariano} G.M. D'Ariano, quant-ph/0208110.
\bibitem{ozawapla} M. Ozawa, Phys. Rev. A {\bf 67}, 042105 (2003).
\bibitem{ozawa} M. Ozawa, quant-ph/0307057.
\bibitem{bohr} N. Bohr, {\it Atomic Physics and Human Knowledge} (Wiley, New 
York, 1958), pp. 32-66.
\bibitem{ak} E. Arthurs and J.L. Kelly, Jr., Bell Syst. Tech. J. {\bf 44}, 725 
(1965).
\bibitem{wootters} W.K. Wootters and W.H. Zurek, Phys. Rev. D {\bf 19}, 473 
(1979).
\bibitem{ag} E. Arthurs and M.S. Goodman, Phys. Rev. Lett. {\bf 60}, 2447 
(1988). 
\bibitem{martens} H. Martens and W. de Muynck, Found. Phys. {\bf 20}, 255 
(1990).
\bibitem{whichway} G. Jaeger, A. Shimony, and L. Vaidman, Phys. Rev. A {\bf 51}, 
54 (1995).
\bibitem{muynck} W.M. de Muynck, Found. Phys. {\bf 30}, 205 (2000).  
\bibitem{trif} A. Trifonov, G. Bj\"{o}rk, and J. S\"{o}derholm, Phys. Rev. Lett. 
{\bf 86}, 4423 (2001).
\bibitem{epr} A. Einstein, B. Podolsky, and N. Rosen, Phys. Rev. {\bf 47}, 777 
(1935).
\bibitem{bohrepr} N. Bohr, Phys. Rev. {\bf 48}, 696 (1935).
\bibitem{wan} K.K. Wan and P.J. Sumner, Phys. Lett. A {\bf 128}, 458 (1988).
\bibitem{holland} P.R. Holland, {\it The Quantum Theory of Motion} (Cambridge 
University Press, UK, 1993), chapter 3.
\bibitem{cohen} L. Cohen, Phys. Lett. A {\bf 212}, 315 (1996).
\bibitem{weak} Y. Aharonov, D.Z. Albert, and L. Vaidman, Phys. Rev. Lett. {\bf 
60}, 1351 (1988).
\bibitem{weakpra} Y. Aharonov and L. Vaidman, Phys. Rev. A {\bf 41}, 11 (1990).
\bibitem{lahti} P. Busch, M. Grabowski, and P.J. Lahti, {\it Operational Quantum 
Physics} (Springer, Berlin, 1995), chapters II, VII.
\bibitem{halleur} M.J.W. Hall, Phys. Rev. A {\bf 64}, 052103 (2001).
\bibitem{weakwiseman} H.M. Wiseman, Phys. Rev. A {\bf 65}, 032111 (2002).
\bibitem{lars1} L.M. Johansen, quant-ph/0308137.
\bibitem{hallfish} M.J.W. Hall, Phys. Rev. A {\bf 62}, 012107 (2000).
\bibitem{luis} A. Luis, Phys. Rev. A {\bf 67}, 064101 (2003).
\bibitem{classical}  Two {\it compatible} (eg, classical) observables $X$ and
$Y$ will always have some joint probability distribution $p(x,y)$.
Minimising the statistical deviation $\langle [X-f(Y)]^2\rangle$ then
leads to the best possible estimate $f(y)=\int dx\,x\,p(x,y)/\int dx\,
p(x,y)$ for $X$ from measurement result $Y=y$, analogous to
Eq.~(\ref{best}) \cite{halleur,luis}.  
Clearly this is always nonnegative for a positive
observable $X$. 
\bibitem{hallalg} M.J.W. Hall, quant-ph/0302007.
\bibitem{yuen} H.P. Yuen and J.H. Shapiro, IEEE Trans. Inf. Theory {\bf IT-26}, 
78 (1980).
\bibitem{shapiro} J.H. Shapiro and S.S. Wagner, IEEE J. Quantum Electron. {\bf 
QE-20}, 803, (1984).
\bibitem{cd} C.M. Caves and P.D. Drummond, Rev. Mod. Phys. {\bf 66}, 481 (1994).
\bibitem{hallfuss} M.J.W. Hall and I.G. Fuss, Quantum Opt. {\bf 3}, 147 (1991).
\bibitem{cover} A. Dembo, T.M. Cover, and J.A. Thomas, IEEE Trans. Inf. Theory 
{\bf IT-37}, 1501 (1991).
\bibitem{bell} J.S. Bell, Physics {\bf 1}, 195 (1964).
\bibitem{shim} A. Shimony, M.A. Horne, and J.F. Clauser, Dialectica {\bf 39}, 97 
(1985).
\bibitem{esp} B. d'Espagnat, {\it Veiled Reality, An Analysis of Present-Day 
Quantum Mechanical Concepts} (Addison-Wesley, Reading, 1995), Appendix 4.
\bibitem{jaynes} E.T. Jaynes, Phys. Rev. {\bf 108}, 171 (1957).
\bibitem{hancock} J.C. Hancock and P.A. Wintz, {\it Signal Detection Theory} 
(McGraw-Hill, New York, 1966), pp. 128-130.
\bibitem{holevo2} A.S. Holevo, {\it Statistical Structure of Quantum Theory}, 
(Springer, Berlin, 2001), Sec.~2.2.4.
\bibitem{hallphaseeprint} M.J.W. Hall, quant-ph/0103072.
\bibitem{ozawacon} M. Ozawa, quant-ph/0310070. 

\end{thebibliography}
\end{document}